# Ten Questions about Emergence

Jochen Fromm, Distributed Systems Group, Kassel University, Germany

**Abstract**

Self-Organization is of growing importance for large distributed computing systems. In these systems, a central control and manual management is exceedingly difficult or even impossible. Emergence is widely recognized as the core principle behind self-organization. Therefore the idea to use both principles to control and organize large-scale distributed systems is very attractive and not so far off.

Yet there are many open questions about emergence and self-organization, ranging from a clear definition and scientific understanding to the possible applications in engineering and technology, including the limitations of both concepts. Self-organizing systems with emergent properties are highly desirable, but also very challenging. We pose ten central questions about emergence, give preliminary answers, and identify four basic limits of self-organization: a size limit, a place limit, a complexity limit and finally a combinatorial limit

## 1. Introduction

Computing systems with very high demand for autonomy, self-organization and self-management consist of many countless components, which are too tiny and small (Ubiquitous and Pervasive Computing), too complex and to numerous (Internet-Applications with thousands of servers), too unpredictable (Mobile Ad-Hoc Networks or MANETs) or too remote (Space probes, for example in the NASA Project Autonomous Nano Technology Swarm named ANTS). Manual management of these distributed systems is difficult, expensive and time-consuming, and often nearly impossible.

It is easy to talk of a new era in computation [Met93], or to invent a new vision or another x-computing paradigm like Amorphous Computing, Organic Computing, Ubiquitous Computing, Grid Computing, Pervasive Computing, Autonomic Computing, Distributed Computing, Recovery Oriented Computing,.... It is much harder to realize such *x-computing* systems – mostly biologically inspired visions – with the old principles and mechanisms, while basic questions are still open and not completely answered. To envision a system running without constant human supervision and intervention is not difficult, to specify how you can achieve this goal is much more difficult, compare for example the glorious and grand vision of autonomic computing [Kep03] with the disillusioning and desperate attempts of realization [Kep05].

First of all, it is important to consider the basic and fundamental questions which are still unsolved despite decades of research. There are many open questions about emergence and self-organization, ranging from a clear definition and understanding to the possibilities and limitations of both concepts. We pose ten central questions about emergence, ranging from science to engineering:

1. Can we understand it?
2. What are the underlying principles?
3. What is the relation between emergence and other concepts like evolution?
4. Are there any necessary requirements like mobility, intelligence or suitable environments?
5. Is it only possible with small, simple and stupid agents?
6. Can we find a kind of calculus based on emergence for Multi-Agent-Systems?
7. How can a globally desired structure or functionality be designed on the basis of interactions between many simple modules?
8. Can we use it? Is it useful?
9. Can we control it? Even if it is not predictable, is it in some way controllable?
10. What are the limitations of emergence and self-organization?

Questions 1-3 are about a clear **definition** (can we understand it?), questions 4-5 about necessary **requirements** (what do we need for it?), questions 6-9 about a possible calculus and **application** (can we model, apply and control it?), and finally question 10 about any existing **limitation** (what are the limits?).

One of the most fundamental and basic questions in the list is number seven. It is an open question since Pattie Maes mentioned it in her classic agent paper from 1994 in the conclusion [Maes94]: "We need a better understanding of the underlying principles. In particular, it is important to understand the mechanisms and limitations of emergent behavior. How can a globally desired structure or functionality be designed on the basis of interactions between many simple modules? What are the conditions and limitations under which the emergent structures are stable, and so on". More than ten years later, these questions are still a matter of research. Scientists are dealing with exactly the same problem, see for example [Yam05]. Therefore a solution is not easy, but would have a very broad-ranging effect.

## 2. Questions

*"In the culture of computer science, an idea that works in one situation is called a hack, an idea that works twice is called a trick, and an idea that works often and pervasively is called a technique"*
Dennis Shasha and Cathy Lazere in [Sha95]

The following ten fundamental and basic questions related to the phenomenon of "emergence" in Multi-Agent Systems are meant to drive further scientific inquiry and will hopefully lead to new engineering techniques. If we want to discover new techniques, models and inspirations, it is certainly useful to look at all major areas where self-organization and "emergence" occur in nature. This means you have not only to restrict yourself on biologically inspired systems, but you should for example consider socially inspired systems and models as well.

The answers to the following questions are preliminary. Some questions are certainly still subject of wild discussions, others may be answered soon. A few will remain open for a while.

### 2.1. Can we understand it? Can we define a comprehensive taxonomy or classification?

Probably yes, we need only to find and reveal the hidden causal connections. In order to understand the phenomena in general, it is possible to start with a crude and coarse taxonomy. We can create a classification according to different feedback types and causal relationships (see [Fro05] at http://arxiv.org/abs/nlin.AO/0506028).

It is difficult to construct a theory or model which describes "emergence" and emergent phenomena, because a property is emergent if it can not be comprehended by the underlying system model - just as something is complex, if it is difficult to describe or no simple description exists.

Complexity means we can not describe the phenomena completely, *because* we do not have a model or description. Emergence means we can not describe the phenomena completely, *although* we have a model and description of local rules and actions. In the first case (complexity) you have no suitable description, in the second case (emergence) you have no suitable model.

### 2.2. What are the underlying principles? How are they related? Is novelty or the need for an observer among them?

It is hard to find a principle of "emergence" or living self-organizing systems which has not been applied to artificial systems. The field of ALife has tried to find and implement these principles for over a decade, from the first conference about ALife held at Los Alamos in 1988 [Lan89] to the ninth international conference on the simulation and synthesis of living systems, ALife IX, in 2004 [Pol04].

The principle or concept which is most closely related to "emergence" is self-organization. The concept of self-organization focuses on the system-environment boundary, if the border between inside and outside or internal and external parts is considered, see figure 1. Although the self-* name emphasizes the autonomy of the system, the self-organizing process is usually not possible in completely isolated and closed systems.

The process of emergence takes places at the boundary between the system and its constituents, if the border between local and global, micro and macro, individual and collective behavior is crossed, see again figure 1. It distinguishes between local, low-level components and global, high-level patterns. Emergence emphasizes the bottom-up process, the appearance of new and novel structures at a higher level, but is not possible without a top-down feedback process.

Self-organization in Multi-Agent Systems (MASs) is closely connected to the phenomenon of "emergence", contrary to other systems. If the terms are used interchangeably in the following, they refer to MAS and agent-based simulations. In **physical systems** with many particles, self-organization is associated with self-organized criticality [Bak96], critical points and phase transitions. In **net systems** with connected nodes, self-organization is realized through rewiring (small-world nets) or "preferential attachment" (scale-free nets). Self-organization in **living systems** is related to self-regeneration, metabolism and autopoiesis.

The concept of "emergence" is less ambiguous and abstract than expression self-organization, because it is a colloquial expression with the meaning of "sudden appearance" as well (at least in English) which can be applied to nearly every system – for instance political, historical, cultural and biological systems. If we consider the colloquial meaning "appearance" in the broad sense, the sudden emergence of something is always possible at a clear boundary or border of something, someone, or some form of system. The list of possible systems is endless.

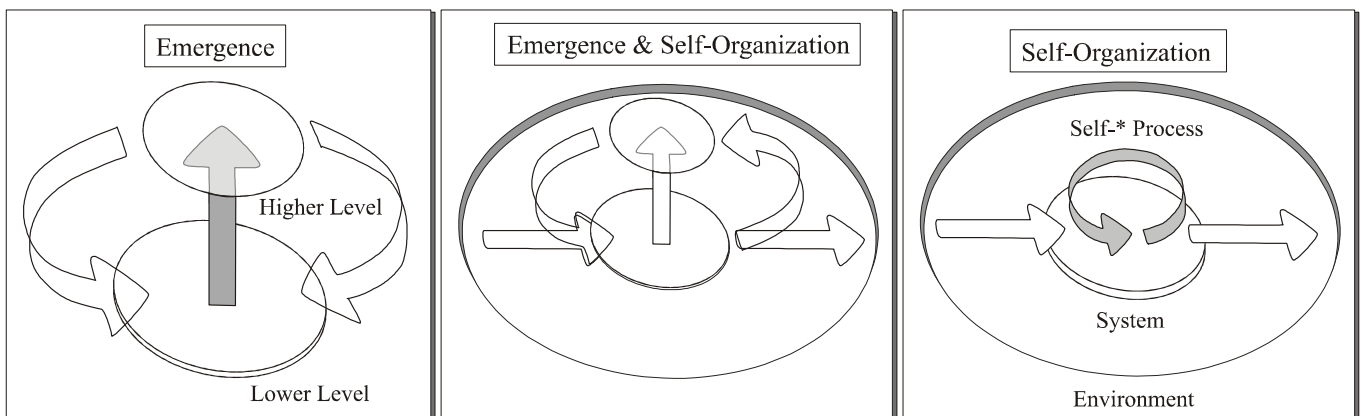

*Fig. 1 Self-Organization and Emergence. From left to right we can distinguish between "Emerging", Self-Perpetuating and Self-Organizing Patterns*

Yet many of these systems can be modelled by Multi-Agent Systems or evolutionary systems. The concept "emergence" applied to evolutionary systems refers to speciation, blockades through fitness barriers, macro-evolution and jumps in complexity through the appearance of species with complex properties. The classic meaning in the context of complex systems – macroscopic order or global patterns arising from local interactions of many microscopic elements – is most clearly visible in MASs.

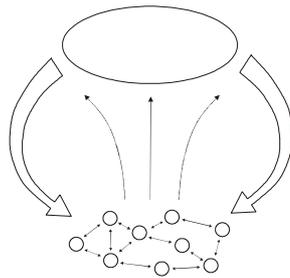

Therefore we focus in the following on the concept of "emergence" in MASs and natural systems which can be modelled by MASs. In MAS, the essential boundary or border is the interface between macroscopic and microscopic level, collective and individual entities, system and agents: the possible roles and role transitions (see question six).

The basic elements and key characteristics of self-organization in Multi-Agent Systems are well-known: a large number of identical agents (redundancy), gradient fields (biochemical scents used in stigmergy and swarm-intelligence in form of pheromones, hormones, and other biochemical substances), multiple interactions (direct or indirect), feedback loops (positive, negative and mixed) and finally some amount of randomness and stochasticity [Bla02, Cam03]:

 - feedback loops
 - causal relationships across different levels
 - right balance between reactive/proactive, context-
   dependence/autonomy and exploitation/exploration

Novelty and surprise are not among the points in the list. Probably you do not need an observer or the notions of novelty and surprise to define emergence.

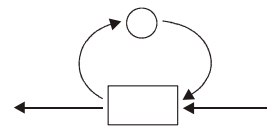

*Feedback loops* are known since the time of Cybernetics in the 1950s and 60s. Already W.R. Ashby noticed the importance of feedback in coupled systems for forms of self-organization which are more than transitions from 'parts separated' to 'parts joined' [Ash62]. He argued that no machine could by itself change its own organization, if the behavior of each part is independent of the other parts' states. He added further that a 'self-organizing' machine is only possible by a machine coupled to another machine.

A feedback loop has two purposes: the feedback is used to control a system, and the feedback signal indicates at the same time the current state of the controlled element. For example emotions in general: they are used to control the body, but the also signal the current state. Pheromone trails control the movement of ants, but they also signal the place of the food and the "foraging" state of the colony.

Emergence is obviously related to hidden causal connections, *causal relationships* across different levels, or whole networks and braids of causal connections (upward, downward, sideward and mixed forms). Any scientific explanation should contain the clarification of causal connections.

Feedback loops across different levels and complicated causal relationships can be found in stigmergy and swarm-intelligence [Cam03]. Both are linked to causal relations across the system-environment boundary. The agents of the system affect the environment, which in turn influences the behavior of the agents. Swarm formation and flocking behavior is more associated with causal relations across the micro-macro boundary, see Fig. 2.

Ozalp Babaoglu et al. [Bab05] have collected a number of design patterns for biologically inspired principles and processes in distributed systems. Besides stigmergy they are mentioning: diffusion (levelling of a concentra-tion gradient) also as a part of reaction-diffusion-systems, replication (epidemic proliferation), chemotaxis (movement towards or away from a concentration gradient). Basically all these processes are related to the modelling of chemical scents and messengers.

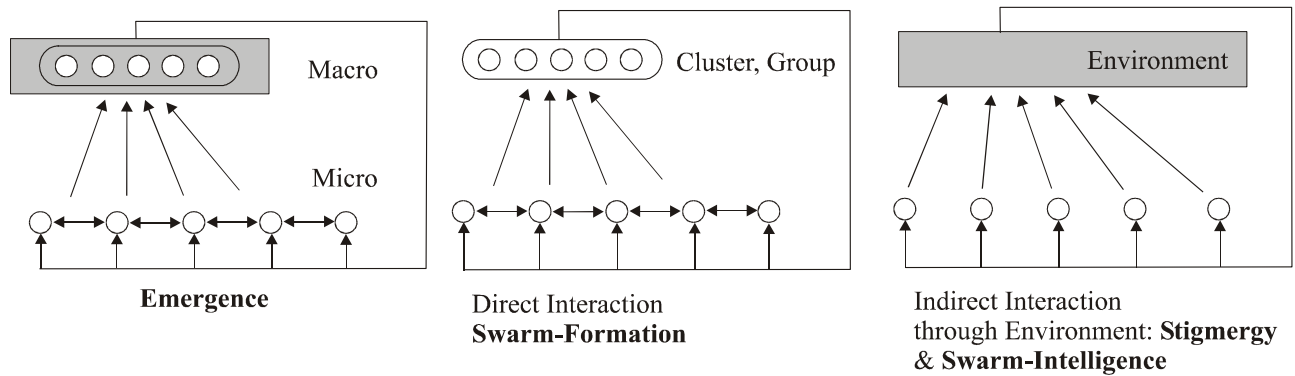

*Fig. 2 Emergence, Swarm-Formation, Swarm-Intelligence*

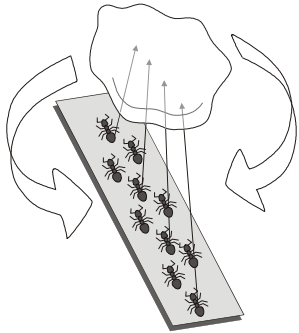

This confirms the observation (see for example [Bon99]) that pheromones are the main principle in the current attempts to imitate natural self-organization in artificial systems, esp. in Multi-Agent Systems. Pheromones are in fact widely used among insects: over 7000 species use them, and over 3000 different substances are known among insects (see for example the pheromone base at http://www.pherobase.com/ ).

### 2.3. What is the relation between emergence and evolution? Are there any processes similar or related to "emergence" in evolution?

*"All things appear and disappear because of the concurrence of causes and conditions. Nothing ever exists entirely alone; everything is in relation to everything else"*                    Buddha

The complexity of nature is deceiving. Self-organization is sometimes identified as the reason for this overwhelming complexity, and "emergence" as the reason for sudden jumps in complexity. Yet evolution is the main reason and the driving force for the complexity and diversity which can be found in nature [Fro04], and neither the concept of self-organization nor the phenomenon of emergence can really replace evolution or natural selection [Cam03].

The evolution of "selfish genes" seems to be responsible itself for all forms of sudden jumps in complexity in history, i.e. the "emergence" of more and more complex species or complex properties in the course of time, and it is also responsible for all forms of "strong emergence", in which whole new evolutionary systems appear (associated with the emergence of a new code which is used to create the new evolutionary system).

Sudden jumps in complexity due to evolution are often related to fitness barriers. There are at least three different ways to cope with fitness barriers:

(1) to wait for a catastrophe, until the barrier is reduced through catastrophic events

(2) to bypass through exaptation: explore a different direction and make a sudden side-leap

(3) to tunnel right through the barrier by borrowing complexity

### 2.4. Are there any necessary requirements for "emergence" like mobility, intelligence or suitable environments? Is emergence a typical property of MAS?

Yes, there are some necessary requirements. If a system shows some form of spatial pattern based on the position of agents and elements, a kind of mobility for agents or movability for elements is certainly needed. If the position of the agents is fixed or arranged in a static grid as a Cellular Automata, then a kind of "tag" or external visible state is required for spatial patterns. Stigmergy and swarm-intelligence are possible only through more or less permanent changes in the environment of the Multi-Agent System:

- agent mobility for spatial patterns if the position of agents varies, visible tags or states if the position is fixed
- changeable environments and some kind of "scent", "pheromone" or evaporating mark for stigmergy and swarm intelligence
- ability to distinguish between groups and individuals for flocking and swarm formation

Emergence in adaptive and evolutionary systems requires special conditions: adaptive systems need a cognitive barrier to enable learning "leaps" and sudden new insights, evolutionary systems need a larger fitness barrier to produce sudden jumps. Both require a complex environment to reach higher levels of complexity.

Emergence is probably not a typical property of every Multi-Agent System (MAS), but it is typical for MAS compared to other traditional forms of programming, since a strong possibility of emergent behavior exists, as Nick Jennings noticed in his classic AOSE article *On Agent-Based Software Engineering*: (a) "the patterns and the outcomes of the interactions are inherently unpredictable" and (b) "predicting the behavior of the overall system based on its constituent components is extremely difficult because of the strong possibility of emergent behavior" [Jen00].

### 2.5. Is "emergence" only possible with simple and stupid agents? Or is it also possible with complex BDI agents?

No, probably simple agents are not necessary to produce emergence of interesting properties. Even complex agents or persons can behave according to simple rules (see e.g. Schelling's segregation model, which describes how slight preference differences can cause segregation in an entire community [Sche78]). But the phenomenon of emergence is better comprehensible and understandable for simple agents, and in fact it is most useful for a large number of small and stupid elements. The context-dependent influence is certainly stronger for simple, stupid and purely reactive agents, and weaker for more complex, intelligent, proactive and goal-directed agents.

This does not mean that emergence is directly opposed to intelligence. Emergent phenomena occur in systems with complex agents as well, and the mind as a whole is itself a favorite example of emergence among philosophers, see [Kim96].

## 2.6. Can we find a kind of calculus for Multi-Agent-Systems?

*"Only by taking infinitesimally small units for observation (the differential of history, that is, the individual tendencies of men) and attaining to the art of integrating them (that is, finding the sum of these infinitesimals) can we hope to arrive at the laws of history."*
War and Peace, by Leo Tolstoy, Book Eleven, Chapter 1

Can we establish universal "laws of history", as Tolstoy said? We probably can not define and predict exactly what a complex system looks like at each time step on the macroscopic level (due to combinatorial explosion, the Butterfly effect, etc). But we can define roughly what the system does: what type of attractors and emergent phenomena do appear – but not when exactly. We can also probably predict how certain types evolve and change roughly – but not in every detail, only on certain time scales.

What the system does can be seen in the roles of the individual agents, which define what the agents do. If the global goal or purpose of an ant colony is to find food in order to survive, then at least at one moment one ant or agent should be busy with the tasks exploration or transport while occupying the roles food-explorer, food-transporter,…. The different agent roles reflect like a broken mirror the overall role, function and purpose of the system. At least the local roles of the agents and the global roles of the system should not completely contradict each other.

Steven Strogatz said in his SYNC Book about complex systems [Stro03]: "I think we may be missing the conceptual equivalent of calculus, a way of seeing the consequences of the myriad interactions that define a complex system". There is probably no way to determine the consequences of myriad interactions in detail. Many microscopic details are insignificant, irrelevant and inconsequential to macroscopic phenomena anyway. What we can do is

* to find the networks and major braids of causal connections (upward, downward, sideward,...): to identify the possible types and forms of emergent phenomena which can appear in the system

* to determine the topology of the interface between the system and its parts

If there is a basic calculus for Multi-Agent Systems (MAS) and Complex Adaptive Systems (CAS) similar to the differential calculus in Mathematics, it should describe changes in the boundary between Agent and System, "topological" changes in agent types, roles and role transitions. The differential calculus in Mathematics is based on differentiation and integration. What has this to do with topology and boundaries? A lot, since integration and differentiation can be found in form of merging/gluing and splitting/cutting as the basic operations in topology which change the boundary, the surface and the topological class. Furthermore, the fundamental theorem of calculus is about changes across the boundary of manifolds[1].

Both operations - differentiation and integration - are also fundamental processes in Object-Oriented Programming (OOP) and Agent-Oriented Programming (AOP): integration in form of aggregation, group formation and composition of objects or agents, and differentiation in form of class, type or role differentiation which is related to specialization, adaptation and inheritance. There are two basic group processes which affect the agent-system boundary: on the one hand we have unifying group-formation/integration processes, and on the other hand separating specialization/role-differentiation processes.

If there is a clear interface or boundary between the system and its constituents, all effects and influences must obviously pass through it. The interface is the common layer or region shared by the microscopic and the microscopic layer, where the influence from the elements to the system and vice versa is strongest. It determines how an element or agent changes its behavior when it becomes the constituent of a system, and what types of behavior are possible while it is a constituent of various parts of the system.

In Multi-Agent Systems, the topology of the interface between system and constituents is determined mainly by the set of roles and possible role transitions. Thus it is not surprising that leading agent researchers agree that roles and organizational concepts are indispensable to model and design MAS, see for example [Fer03,Zam03].

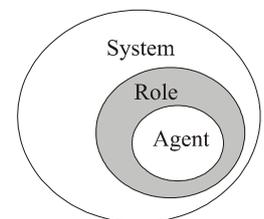

How a methodology, framework or a general calculus would look like in detail becomes clearer after the next question. Natural self-organizing systems can be simulated well, but the engineering and goal-directed design of general self-organizing systems with emergent properties is a core problem, because controlling behavior is an inherent and intrinsic problem in these systems. It is difficult to construct a specific system if you neither can understand nor predict and control it completely.

---

[1] Stokes theorem says roughly that the stream through a boundary of a manifold is equal to the field change in the enclosed volume and can be considered as a generalisation of the fundamental theorem of calculus. Differentiation ("the slope of the curve") describes what makes a function special, how a function grows and shrinks at a certain point, how much units cross through the "boundary". Integration ("the area under the curve") describes accumulation, how much has changed in a certain region or interval limited by a particular "boundary".

## 2.7. How can a globally desired structure or functionality be designed on the basis of interactions between many simple modules? Can we solve the micro-macro link problem?

*"blind deduction from constituent laws can never bulldoze its way through the jungle of complexity generated by large-scale composition"*
Sunny Y. Auyang, [Auy98]

This is the central question [Mae94] and the most difficult one [Ser04]. How can we generate complex global behavior from simple local actions? As Robert Axelrod said [Axe97], how do you use simple local rules to generate higher levels of organization from elementary actors? How can we *design* local behavior so that a certain global behavior emerges? How can we *engineer* self-organizing systems? Giovanna Di Marzo Serugendo argued "this is difficult, because the global goal is not predictable as the sum or a function of the local goals" [Ser03].

The engineering and designing Multi-Agent Systems is hard. Katia P. Sycara says [Syc98]: "Designing and Building agent systems is difficult. They have all the problems associated with building traditional distributed, concurrent systems and have the additional difficulties that arise from having flexible and sophisticated interactions between autonomous problem-solving components."

To engineer self-organizing systems is as difficult as adding self-organization to traditional software engineering [Ser04]. Adding self-organization to traditional applications means currently the addition and consideration of self-* properties [BJM05,Kep03]. This difficulty in engineering and design is not accidental or incidental, it is an inherent disadvantage of Multi-Agent Systems (MASs) with many autonomous agents. Compared to traditional applications and software systems, we can list the following advantages and disadvantages of self-organizing systems and MAS:

Natural advantages, positive properties
- robustness, adaptiveness, fault-tolerance
- scalability, concurrency
- adaptability, flexibility, low brittleness

Negative properties and drawbacks
- low predictability and understandability
- controlling (emergent) behavior is difficult
- engineering and design is hard
- accidents and errors possible
- restricted reliability for computational purposes

Nevertheless Mamei and Zambonelli say that "there is a great need for general, widely applicable engineering methodologies, middleware and APIs to embed support and control self-organization in MAS" [Ser04]. Thus it would be highly desirable to construct systems without the inherent disadvantages and drawbacks.

**Methodologies**

The huge number of diverse Agent-Oriented Software Engineering (AOSE) Methodologies proves that there is a big need for such a methodology, and yet at the same time the existence of a major obstacle: engineering (in form of software applications) and autonomy (in form of autonomous agents) don't seem to fit well together. A large number of methodologies claims to solve this problem:

- French, **ADELFE** (Atelier de Développment de Logiciels à Fonctionnalité Emergente) from Bernon and Gleizes [Ber03]
- British, **GAIA** 1st version from Wooldridge, Jennings & Kinny [Woo00], refined version from Zambonelli, Jennings & Wooldridge [Zam03]
- Italian/Canadian, **Tropos** 1st Version from Bresciani, Perini, Giorgini, Giunchiglia, Mylopoulos, [Bre01], refined version from Kolp, Giorgini and Mylopoulos [Kol02]
- German, **MASSIVE** (Multi Agent Systems Iterative View Engineering) Lind [Lin01]
- American, **MaSE** (Multiagent Systems Engineering) from Wood and DeLoach [DeLo99]
- Australian, **Prometheus** from Padgham and Winikoff [Pad02]

ADELFE is one of the few methodologies which claim to support emergent behavior, since "emergent functionality" is already part of the name (Atelier de Développment de Logiciels à Fonctionnalité Emergente). Yet it remains to be shown that you can really construct interesting new forms of emergent behavior with Adelfe.

MASSIVE and Prometheus are nearly the only methodologies which mention the need for an iterative process. Especially MASSIVE emphasizes the importance of "Round-trip Engineering" and "Iterative Enhancement". A stepwise refinement and iteration is probably essential to construct systems with emergent properties.

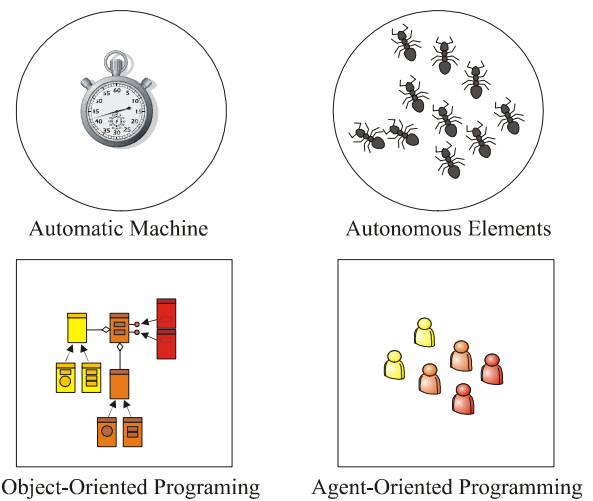

*Fig. 3 Objects and Agents, OOP and AOP*

The importance of roles, organizations and organizational structures is emphasized by nearly every approach and methodology. On the conceptual level, a collection of software-objects works like a machine, whereas a collection of agents is more like a society or community. Organizations are the natural way to organize a society. Yet most methodologies like GAIA require to specify the overall structure of the MAS organization and to define appropriate organizational abstractions. Perhaps this does not completely contradict the idea of self-organization, but the need for detailed specification of the organizational structure at design-time does not leave much room for self-organization.

Most methodologies have a strong similarity to existing object-oriented methodologies, for example Adelfe, Gaia and Tropos are all based on the three core steps (1) Definition of Requirements (2) Analysis and (3) Design. They rely on the analysis and design of a fixed and static organization, contrary to dynamic natural organizations which evolve and grow, and sometimes even organize themselves. The concepts "self-organization" or "emergence" are seldom used in AOSE methodologies. None of all these different methodologies seems to be a satisfactory solution of the micro-macro link (MML) problem. The existing AOSE (Agent Oriented Software Engineering) methodologies are not suitable to solve the ESOA (Engineering of Self-Organizing Applications) problem, although the word AOSE is the expression ESOA read backwards.

Is it possible that there are so many different methodologies, because engineering and autonomy, objects and agents, object-oriented programming (OOP) and agent-oriented programming (AOP) are like oil and water? They do not seem to fit well together on a conceptual level (although agents are of course on a practical level often implemented in object-oriented languages). Is the engineering of self-organizing systems possible at all?

We can give only a preliminary and coarse answer to the general question. As Jonathan Rauch argued [Rau02], we will probably not be able to foresee the future of each possible MAS in every detail, but we might learn to anticipate the kinds of events that lie ahead. Even if we manage to create a methodology, framework or a general calculus for the engineering of self-organizing systems, some properties of the system will possibly remain uncertain.

The question of the Micro-Macro Link (MML) is also a core problem in DAI and Sociology [Schi00]. Can we find a reliable MML from local to global behavior and back? The central task of the MML problem (to link the behavior of the constituents and the system) is as already mentioned in the answer of the last question to find the networks and major braids of causal connections and to determine the topology of the interface between the system and its parts. How do we do this?

**A Two-Way Approach to the MML**

As Conte and Castelfranchi have argued, [Con95] the micro-macro link (MML) problem probably needs a two-way or two-phase approach to find the necessary micro-macro connections, including a bottom-up and a top-down process. The way up determines how individual actions are combined and aggregated to collective behavior, the way down defines how collective forces influence and constrain individual actions.

We can only generate emergent properties in a goal-directed, straightforward way if we look at the microscopic level and the macroscopic level (for local *and* global patterns, properties and behaviors), examine causal dependencies across different scale and levels, and if we consider the congregation and composition of elements as well as their possible interactions and relations. A complex system can only be understood in terms of its parts *and* the interactions between them, if we consider static *and* dynamic aspects.

In other words we need a combination of top-down and bottom-up approach, which considers all sides: static parts and dynamic interactions between them, together with the macroscopic states of the system and the microscopic states of the constituents. Sunny Y. Auyang proposes a method named "synthetic microanalysis" which claims to combine synthesis and analysis, composition and decomposition, a bottom-up and a top-down view, and finally micro- and macrodescriptions [Auy98]. She describes the idea vividly in chapter 2 of her interesting book, but unfortunately she does not say how her approach works for MAS exactly.

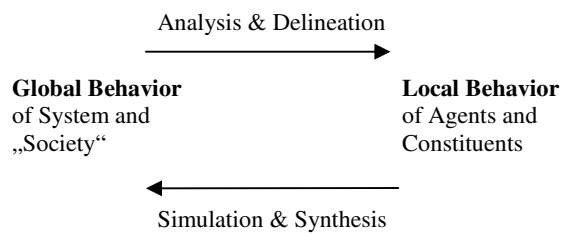

*Fig. 4 Synthetic Microanalysis*

The general idea is a "bottom-up deduction guided by a top-down view". The detailed process is not yet clear for MAS in general, but you roughly have to delineate groups of "microstates" according to causal related macroscopic criteria (by partitioning the microstate space, for instance through selection of all elements with a certain property or role related to some macroscopic structure). In other words you try [Auy98] "to cast out the net of macroconcepts to fish for the microinformation relevant to the explanation of macrophenomena" (p.56). If you "make a round trip from the whole to its parts and back" [Auy98], you can use the desired global macroscopic phenomena to design suitable local properties and interactions.

The bottom-up approach alone is successful only for small and simple systems like 1-dim Cellular Automata, where you can enumerate all possible systems. For large systems the amount of possibilities and number of configurations grows so large (or even "explodes") that the goal gets lost or the thicket of microscopic details becomes impenetrable. To quote Auyang [Auy98] again: "blind deduction from constituent laws can never bulldoze its way through the jungle of complexity generated by large-scale composition" (p.6).

The macroscopic view is useful and necessary to delineate possible configurations, to identify composite subsystems on medium and large scales, to set goals for microscopic simulations and finally to prevent scientists "from losing sight of desired macroscopic phenomena when they are immersed in analytic details" [Auy98].

**Iterations and Refinements**

One round trip from the whole to its parts and back is probably not enough to generate complex self-organizing systems with emergent phenomena. If the two-way method of "synthetic microanalysis" works at all, you will certainly need some iterations and a number of stepwise refinements until the method converges to a suitable solution.

Before each iteration, it is important to identify and refine suitable subsystems, basic compounds and essential phenomena on the macroscopic level, which are big and frequent enough to be typical or characteristic of the system, but small and regular enough to be explained well by a set of microscopic processes. Many macroscopic descriptions are only an approximation, idealization and simplification of real processes. Often they must rely on probabilistic, stochastic or statistical concepts, because there is too little information.

In the first top-down phase towards the bottom level, we must find the significant, relevant and salient properties, events and interactions, especially the crucial events responsible for butterfly effects, avalanches and cascades. We seek the concrete, precise and deterministic realization of abstract concepts. Many microscopic details are insignificant, irrelevant and inconsequential to macroscopic phenomena, there is too much information. In the second bottom-up phase towards the top level, you have to compare the results of the synthesis and simulation which the desired structure.

In a typical iteration of "synthetic microanalysis", you start from the "top" and work your way down to the micro-level, constructing agent roles and interaction rules in just the way necessary to generate the behavior observed on "top". This procedure can be iterated by stepwise refinement of agents and their interactions, which should include necessary changes in the environment, until the desired function is achieved.

In the next round, you start start again from the global structure or macroscopic pattern, and try to refine the possible underlying microstates and micromechanisms. Could these states and mechanisms lead to the desired large-scale structure? What kind of coordination, conflict-resolution and local guidance is needed additionally? What kind of roles and role-transitions are possible?

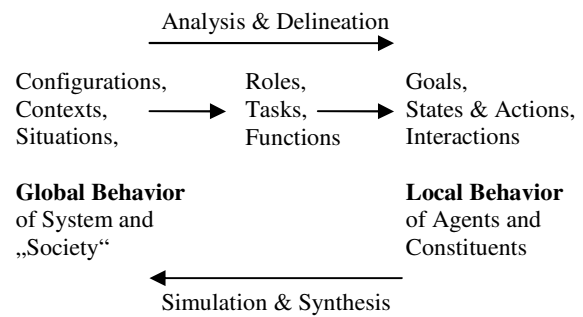

*Fig. 5 Synthetic Microanalysis in Detail*

Thus you would proceed roughly like this while trying to determine possible states, roles and role transitions:

### Phase 1. Analysis and Delineation

What macroscopic patterns, configurations, situations and contexts are possible in principle? From the answers you can try to delineate what roles, behaviors, local states and local interactions are roughly possible or necessary:

a) What roles and local behavior are possible? Try to determine and deduce local behavior from global behavior, identify possible roles and role *transitions*.

b) What states possible? Determine and define local properties from global properties.

c) What kind of local communication and coordination mechanisms possible? Determine tolerable conflicts and inconsistencies.

### Phase 2. Synthesis and Simulation

Is the desired global behavior achievable with the set of roles and role transitions? In the second phase, you work your way up to the top through comprehensive simulations.

Simulation is the only major way up from the bottom to the top. As Giovanna Di Marzo Serugendo says "the verification task turns out to be an arduous exercise, if not realized through simulation" [Ser03].

Sometimes emergence is even defined through simulation, for instance in the following way: a macrostate is weakly emergent if it can be derived from microstates and microdynamics but only by simulation [Bed97].

The way up is much simpler than the way down and requires mainly simulations. Since these simulations can be quite time consuming, it can be slower than the top-down process. In mathematical calculus, the situation is quite similar: many integrals can only be solved and determined numerical by numeric calculations, whereas differentiation is much easier and requires often only sophisticated analysis and analytic techniques.

There are more similarities: the fundamental theorem of calculus also connects the purely algebraic indefinite integral and the purely analytic (or geometric) definite integral[2]. Likewise, a method of synthetic microanalysis should combine simulations (preferably bottom-up) and "analytic" (preferably top-down) considerations.

If a general methodology to engineer self-organizing Multi-Agent Systems exists, it will probably look similar: it will consist of a two-way approach with two phases like synthetic microanalysis, it will contain the emphasis of "emergence" from ADELFE, the "Round-trip Engineering" and "Iterative Enhancement" known from MASSIVE, the concepts of roles and organizational structures from GAIA, and the highlighting of goals and multi-agent patterns from TROPOS.

## 2.8. Can we use it? Is "emergence" one reason why agents are not widely used or why agents could be a new successful "programming paradigm"?

We can only use it if we can understand and control the phenomenon. This means we can find a useful methodology, framework or a general calculus, see the preceding central questions 6 and 7. "Emergence" is probably not an essential reason why agents are not widely used, but certainly a factor which makes the engineering and design of MAS very difficult.

The lack of suitable environments and applications is the most important reason why agents are not widely used in software engineering. If a methodology or calculus for the engineering of self-organizing systems with emergent properties can be found, useful applications possibly become more numerous. Such a methodology for the engineering of self-organizing systems could increase the success of agents and Multi-Agent Systems.

## 2.9. Can we control it? Even if "emergence" is not predictable, is it in some way controllable?

There is a strong possibility of emergent properties in MAS (see Question 4), and systems with these properties are unpredictable and therefore hard to control. Emergence is inherently unpredictable.

The situation becomes worse in large, distributed systems. The more distributed and evolutionary a computing system becomes, the more it slips out of control [Kel03], and the bigger the need for autonomy, self-organization, self-managing and other "self"-properties. If we manage to build an engineer system with these properties in a goal-directed way, and if we can anticipate roughly what kind of events can occur, and what types of emergent properties can arise, the system is much more controllable than a system where we can observe and identify no regularities at all.

---

[2] http://mathworld.wolfram.com/DefiniteIntegral.html

Even if "emergence" is not predictable completely, we can identify possible types and forms of emergence by theoretical considerations and anticipate concrete emergent phenomena through practical simulations (which are, as we have argued above, necessary in any methodology for the engineering of self-organizing systems).

Thus the answer depends again on the answers for the central questions 6 and 7. A methodology for the engineering of self-organizing systems would by definition be controllable (otherwise it would not deserve its name), although the system may not be predictable in every detail.

## 2.10. What are the limitations of emergence and self-organization?

It is important to remember that the often misused concept self-organization is the exception, not the rule. Self-organization can certainly be observed in many animal societies and social insects, examples are wasp nests, termite mounds, honey bee combs, ant trails, fish schooling, … [Cam03].

While "self-organization" is frequently seen as the cause for complexity in nature (since nobody "organizes" nature), "emergence" is sometimes mistaken for the origin of jumps in complexity. Yet neither self-organization nor emergence is responsible for overwhelming complexity heights or sudden changes in complexity. Both concepts are limited through the following four limitations:

- **size limit**: the concepts are most useful for a large number of small elements and less useful for small number of very large elements
- **place limit**: the concepts are not possible in every system, only in some, for example in open systems or at the "edge of chaos"
- **complexity limit**: the corresponding phenomena are not arbitrary complex (really complex forms are usually the result of pervasive evolution)
- **combinatorial limit**: if the number of combinatorial possibilities and possible combinations gets too large or explodes (as it can be observed in cases of "strong emergence"), then evolution forms closed entities like cells or organs, and finally invents a new code

### - Size Limit: Small Systems

The concept of self-organization is so vague that it can be applied to situations of any size and complexity, even to the world or universe as a whole [Jan80]. The related concept "emergence" can be used likewise in order to explain *that* the human mind arises from the interactions of myriads of neurons. But in both cases you are not able to explain *how* the human mind arises exactly from the huge neural collective, or how the living beings on earth organize themselves precisely. All you can say is that the earth as a whole consumes a certain amount of energy and produces a huge amount of entropy [Kle05].

The emergence of the mind is in fact a very favorite concept in philosophy [Kim96], but like the philosophical name supervenience an empty or shallow term for a very complex process which expresses our inability and helplessness to describe the involved processes in detail. The emergence of ancient cultures is also a complex process, which is not only related to social questions as division of labor, but also to a lot of other essential factors like taxes, laws, languages and writing systems, among many other things (and which is involved in the appearance of political, economic and other important systems as well).

It is obvious that you need suitable abstractions and models to explain such complex processes. We have the choice between a small number of large elements, and a large number of small elements: 'swarm' systems. The concepts "self-organization" and "emergence" can be applied to both, but the larger the elements, the less probable is a precise explanation of the involved processes. If the elements refer to large, complex entities, they must be abstract enough to say anything useful.

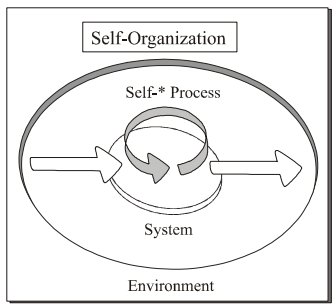

Self-organization occurs traditionally at the boundary between system and environment: to be a useful concept, it requires that the size of the system is not too large. Self-organization therefore refers often to situations where a large number of small elements or a swarm of stupid individuals with limited cognitive abilities are involved and where centralized control is unable to work.

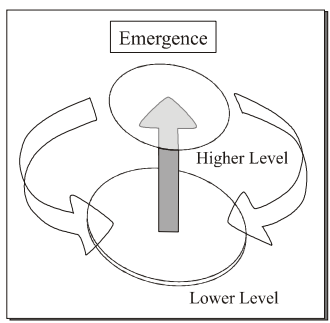

Emergence occurs at the micro-macro boundary: to be a useful concept, it requires that the distance between microscopic and macroscopic level is not too large. It works best in situations where the distance between macroscopic and microscopic layer remains small, or in other words if there are not too many intermediate layers, sheets and levels involved in the process.

Large biological life-forms and systems do not assemble themselves autonomously from many independent elements. They grow under the detailed control and command of selfish genes. They have many layers (molecule, organelle, cell, tissue, organ, organism) and may contain several subsystems like the autonomic nervous system, the limbic system (that produces emotions), the endocrine system (that produces hormones), the immune system (that produces antibodies), the metabolic system,…. which have the task to control and regulate certain system parameters. Yet this is no self-organization in the sense of organization without organizer, it is more a form of self-management or autonomic computing [Kep03].

### - Place Limit: Open Systems

The idea of self-organization challenges the idea of ever-decreasing order based on the second law of thermodynamics. However, the two do not need to be in contradiction: it is possible for a system to reduce its entropy by transferring it to its environment. In open systems, it is the flow of matter and energy through the system that allows the system to self-organize, to import order and organization and to export entropy and disorder to the environment. Therefore open systems can exhibit self-organization, while isolated systems cannot decrease their entropy. Biological systems are open systems feeding from the environment and dumping waste and garbage into it.

The second law of thermodynamics corresponds to our intuitive expectation that order does not increase spontaneously. Left to itself, a thing does not become more organized. Your desktop does not organize itself, your bookshelf does not order itself, and your coffee-cup does not extract heat from the room to heat itself to the right temperature. Left to themselves, systems become more messy, orderless and disorganized. Self-organization is possible in open systems, but it is the exception. Order does not "pop up" for free at every corner. As C.R. Shalizi says in his notebooks on self-organization, "when we encounter a very high degree of order, or an increase in order" we expect that "something, someone, or at least some peculiar thing, is responsible"[3]. This expectation is right. As already said, Self-Organization is the exception. It does not occur everywhere for free.

Typical examples for a closed system are cellular automata. They are closed models and systems (an exception are the extended open models used by Per Bak to simulate sand piles and earthquakes [Bak96]), and the complexity which can arise in them is limited, because there is no inflow of information or energy from the outside. A closed system can neither extract order from the environment nor export disorder to it.

Self-organization in the sense of phase transitions or self-organized criticality [Bak96] often occurs addition-ally only at "the edge of chaos" or at critical points. The emergence of avalanches, earthquakes, cascades and general catastrophes at these points can be described by a power-law: large events are rare, small events happen frequently.

---
[3] http://cscs.umich.edu/~crshalizi/notebooks/self-organization.html

Life always builds on other life, and complex adaptive organisms in nature appear in regions with high complexity. The more complex a system is in nature, the more local it seems to be. The general "emergence" and appearance of more and more complex composite objects is only possible through an increased localization and confinement to a limited space (see chapter 2.1 of [Fro04]). Thus, "emergence" is mostly possible only in special places, see also the requirements in question four: in open or self-organizing systems, in systems with suitable conditions and environments, at critical points, at "the edge of chaos" and in regions with high complexity.

Because self-organization is like organization, pattern and order a vague and ambiguous idea, the concept of "self-organization" or the phrase "order for free" easily lead on the wrong track. In most cases where systems are amazingly complex there is also some other basic process involved, often evolution or metabolism. Quite frequently such systems are the result of a very long evolutionary process. Metabolism is the self-regenerating process which maintains the continuous growth and regeneration of biological bodies. It is based on the principle built-up (anabolism) through breakdown (catabolism).

### - Complexity Limit: Evolution

Intricate organization is the hallmark of a a complex system - for example an economy, a weather system, an immune system, a liquid or a brain. Although the elements and elementary components are simple - for example atoms or agents - the organization and organized interaction makes the system complex. Such a system is what it is and does what it does because of the way in which its constituent parts are organized and not because of what they are. It can only be understood in terms of its parts and the interactions between them.

Since in many natural systems no central organizer is visible, the systems often appear to be the result of self-organization. But many of these systems are also subject to and result of evolution, the most basic principle of biology. In fact evolution and natural selection act on nearly all living systems, and in many cases their effect is much stronger than short-lived forms of self-organization.

In figure 6 we can see the different influences of self-organization and evolution for the different types and forms of emergence, according to the classification in [Fro05]. With increasing complexity (from nominal or simple emergence in Type I to weak emergence in Type II, multiple emergence with many feedbacks in Type III and strong emergence in Type IV), the influence of evolution becomes stronger, see also question 3.

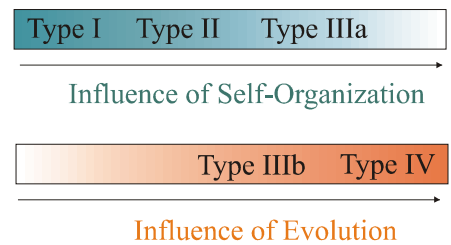

*Fig.6 Self-Organization vs. Evolution in different "Emergence" types*

In nature, we find numerous examples of simple two-dimensional ("striped", "meshed" or "mottled") and three-dimensional ("porous", "clustered") patterns, and simple structures as piles, stripes and dunes, which can be explained well by reaction-diffusion models, autocatalysis and self-organized criticality [Bak96, Cam03]. Yet whenever we consider really complex forms and complicated structures, we hit on traces of evolutionary influences, since nearly every complex life-form is subject to evolution.

### - Combinatorial Limits: Codes

With initiatives and visions as organic computing and autonomic computing, scientists and engineers try to construct biologically inspired systems. The hope is that by learning from organic systems, we can discover and apply new forms of distributed and decentralized organization. Yet there is probably no mysterious unknown principle of self-organization. Many elementary tools such as pheromones in ant colonies and waggle dances in honey bee swarms are well known, and they are a concrete form of language.

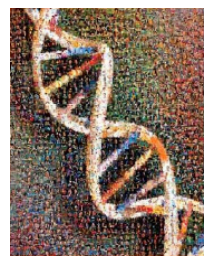
Even nature and its universal tool, evolution, seem to avoid systems with very high complexity. If the number of combinatorial possibilities and possible combinations gets to large or explodes (as it can be observed in cases of "strong emergence"), then evolution forms closed entities like cells, cell nuclei or organs, and finally invents a new code (genetic code in form of DNA or memetic code, i.e. normal language). The emergence and appearance of a new code seems to mark the limits of self-organization.

A code is a set of rules, instructions and symbols, which can store information and messages. It can be used to control and organize processes. With genetic and memetic codes, evolution created a model in form of a serial language or linear string, which is a new layer of abstraction and at the same time a tool to describe and control activities and processes. This code can also be a binary code (0/1), a dynamic code as a waggle dance used by honey bees, or a pheromone code in form of a gradient field.

Useful information stored in a code can be used, re-used, altered and changed later on. A code reduces dependency from the current moment and increases independence and autonomy. For evolution, a new code means the gateway to a new evolutionary system.

Human engineers have designed and invented a lot of codes: the binary code independent from electronic devices, assembler code independent from specific binary coding of instructions, high level programming languages independent from specific processor instruction sets, and finally platform independent Java bytecode and the code of the Common Intermediate Language (CIL) from .NET.

Thus new sophisticated forms of emergence and self-organization in distributed systems and Multi-Agent Systems potentially require the invention of new, partially closed and isolated building blocks, and the use of a new "code" to specify and store the behavior of these basic blocks.

## 3. Conclusion

We have posed ten central questions about "emergence" in Multi-Agent Systems. The answers are preliminary, influenced by personal biases and subjective judgment. Nevertheless they may help to drive further scientific inquiry and will hopefully lead to new engineering techniques. Some questions are certainly still subject of wild and controversial discussions, others may be answered soon. A few will remain open for a while.

The central question is the problem of the Micro-Macro Link (MML). In order to solve the MML problem and to understand "emergence" in a complex system, you need to find the networks and major braids of causal connections (upward, downward, sideward,...), to identify the possible types and forms of emergent phenomena which can appear in the system, and finally to determine the topology of the interface between the system and its parts. In Multi-Agent Systems (MASs), the topology of the interface between system and constituents is determined mainly by the set of roles and possible role transitions.

The design and engineering of self-organizing MAS is difficult. This difficulty is not accidental or incidental, it seems to an intrinsic and inherent problem related to MAS and autonomous agents in general. Engineering and autonomy interfere with each other, and other factors like evolution and "emergence" make the control of such systems difficult. Existing methodologies for Agent Oriented Software Engineering (AOSE) require the definition of static organizations, roles and organizational structures, completely contrary to self-organizing systems with emergent properties in nature, which grow, evolve and even organize themselves. Therefore the existing AOSE methodologies are not suitable to solve the ESOA (Engineering of Self-Organizing Applications) problem.

If a general solution of the MML and ESOA problems in is possible, then it will probably a kind of iterative two-way or two-phase approach (similar to Auyang's synthetic microanalysis) with stepwise refinements. Simulations are essential for the verification of a system with unpredictable emergent properties, but not enough to find the right way through the jungle of complexity generated by large-scale composition. The macroscopic view is also necessary to delineate possible states and configurations, to identify composite sub-systems on medium and large scales, and to set goals for microscopic simulations.

We delineate how such a general two-way method of synthetic microanalysis for MAS would look like, compare it to existing AOSE methodologies, and draw analogies to the traditional differential calculus in mathematics.

Although such a methodology would be a very powerful tool, and the concept of self-organization and emergence are fascinating subjects, we should also be aware of the limitations. There is no magic form of self-organization or emergence. Things usually do not organize themselves, and they usually do not appear from nowhere. It is important to remember that the often misused concept self-organization is the exception, not the rule. While "self-organization" is frequently seen as the cause for complexity in nature, "emergence" is sometimes mistaken for the origin of jumps in complexity.

Yet neither self-organization nor "emergence" is really responsible for overwhelming complexity heights or sudden changes in complexity. Both concepts are often confused with evolution and are limited at least through the following four limitations: a size limit, a place limit, a complexity limit and finally a combinatorial limit. Whenever we consider really complex forms and complicated structures, we hit on traces of evolutionary influences, since nearly every complex life-form is subject to evolution.